\documentstyle[twoside,fleqn,espcrc2,epsf]{article}

\title{Massless Majorana fermion on the domain wall}

\author{T. Hotta\address{Institute of Physics, University of Tokyo,
    Komaba, Meguro-ku, Tokyo 153, Japan}%
  ,
  T. Izubuchi\address{Institute of Physics, University of Tsukuba,
    Tsukuba, Ibaraki 305, Japan}
  and 
  J. Nishimura\address{Department of Physics, Nagoya University,
    Chikusa-ku, Nagoya 464-01, Japan}}

\begin{document}

\begin{abstract}
  We study the domain-wall formalism with additional Majorana mass
  term for the unwanted zero mode, which has recently been proposed
  for lattice construction of 4D ${\cal N}=1$ super Yang-Mills theory
  without fine-tuning. Switching off the gauge field, we study the
  dispersion relation of the energy eigenstates numerically, and find
  that the method works for reasonable values of Majorana mass.
  We point out, however, that a problem arises for too large Majorana
  mass, which can be understood in terms of the seesaw mechanism.
\end{abstract}

\maketitle

\section{Introduction}

Supersymmetry is difficult to realize on the lattice.
This is not so surprising since the lattice regularization breaks the
translational invariance, which forms a subgroup of the
supersymmetry.
As the translational invariance is restored in the continuum limit, we
can restore supersymmetry in the continuum limit.
But the price we have to pay for the latter is that we need
fine-tuning in general.
For 4D ${\cal N}=1$ supersymmetric Yang-Mills (SYM) theory, one can
use the Wilson-Majorana fermion for the gaugino and recover
supersymmetry in the continuum limit by fine-tuning the hopping
parameter to the chiral limit \cite{CV}.
Some numerical works have been started along this line \cite{num}.
Fine-tuning is a hard task, however, as is known in the numerical
studies of the chiral limit in QCD, and a method without fine-tuning
is highly desired. 

The overlap formalism \cite{NN} can be used for this purpose, since it
preserves exact chiral symmetry on the lattice.
The problem here is that the formalism is not suitable for numerical
simulation as it stands.
A practical proposal made by Ref. \cite{nishimura} is to use the
domain-wall formalism \cite{kaplan,FS}, and to decouple the unwanted
zero mode by adding Majorana mass term for it.

In this article, we examine whether this approach really works when
the gauge field is switched off as a first step.
We diagonalize the Hamiltonian of the system numerically and study the
dispersion relation of the energy eigenstates for various additional
Majorana mass.

\section{The model}

The action of the model consists of two parts:
\begin{equation}
  \label{action}
  S = S_0 + S_{\rm mass}.
\end{equation}
$S_0$ is given by
\begin{eqnarray}
  \label{action0}
  S_0 & = &\displaystyle \sum_{s, t} \{ \bar{\xi}_s
    \delta_{s t} \sigma_\mu \partial_\mu \xi_t + \bar{\eta}_s
    \delta_{s t} \bar{\sigma}_\mu \partial_\mu \eta_t \nonumber \\
    & & + \bar{\xi}_s {\cal M} \eta_t +\bar{\eta}_s {\cal M}
    ^\dagger \xi_t \} , 
\end{eqnarray}
where ${\cal M} = \delta_{s+1 \ t} - \delta_{s t} \left ( 1 - M -
\frac{1}{2} \triangle \right)$.
With this action, one obtains a right-handed massess Weyl fermion and
a left-handed one localized at the boundaries of the fifth direction
$s = 0, N_s$, respectively.
In the $N_s \rightarrow \infty$ limit, the chiral symmetry is exact
and we end up with one massless Dirac fermion \cite{FS}.
For finite $N_s$, the chiral symmetry is broken but this breaking
vanishes exponentially with increasing $N_s$ \cite{Vranas}.

When we consider 4D ${\cal N}=1$ SYM theory, the fermion is in the
adjoint representation, which is real, and hence the Weyl fermions can
be viewed as Majorana fermions, one of which has to be decoupled,
while the other kept exactly massless.
Let us identify the zero mode in $\xi$ as the massless Majorana
fermion we want, namely the gaugino.
In order to decouple the unwanted zero mode in $\eta$, we introduce
the additional term $S_{\rm mass}$ to the action.
Note that since the fermions are in a real representation, we can
introduce Majorana mass term for each Weyl fermion, $\xi$, $\eta$,
independently, without violating the gauge symmetry.
There is a variety of choice for the $S_{\rm mass}$, but as the
simplest one, we restrict ourselves to the Majorana mass term for the
$\eta$ localized at the boundary $s = N_s$ \cite{nishimura}:
\begin{equation}
  \label{majorana}
  S_{\rm mass} = \displaystyle m \sum_s \left. \left ( \eta^T_s \sigma_2
  \eta_s + \bar{\eta}_s \sigma_2 \bar{\eta}^T_s \right ) \right|_{s =
    N_s}.
\end{equation}

\section{The dispersion relation}
\label{hamiltonian}

We derive the Hamiltonian of the system from the action and
diagonalize it numerically.
Since we have switched off the gauge field, the system is
translationally invariant, and therefore we can partially diagonalize
the Hamiltonian by working in the momentum basis.
We calculate numerically the energy eigenvalues for fixed
three-dimensional momentum $\bf p$.

\begin{figure}[htbp]
  \begin{center}
    \leavevmode
    \epsfysize=6.5cm
    \epsffile{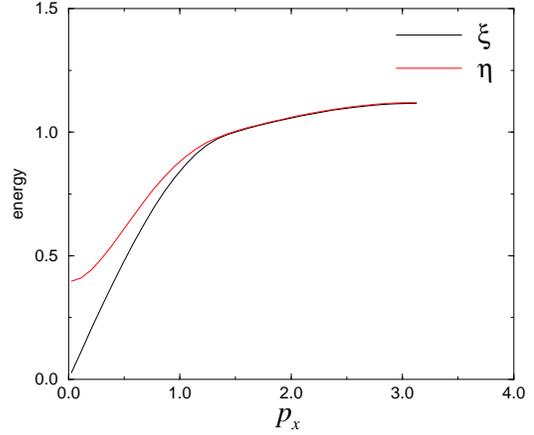}
    \caption{The dispersion relation of $\xi$ and $\eta$ for $m =
      0.2$}
    \label{fig:dispersion1}
  \end{center}
\end{figure}

For moderate Majorana mass we have one massless Weyl fermion localized
at $s = 0$ as expected.
Figure~\ref{fig:dispersion1} shows the energy of $\xi$ and $\eta$ as a 
function of $p_x$ where $p_y = p_z = 0$ when the Majorana mass is
$0.2$.
One can see that the $\xi$ has a linear dispersion relation, while the
$\eta$ has a mass gap.

\begin{figure}[htbp]
  \begin{center}
    \leavevmode
    \epsfysize=6.5cm
    \epsffile{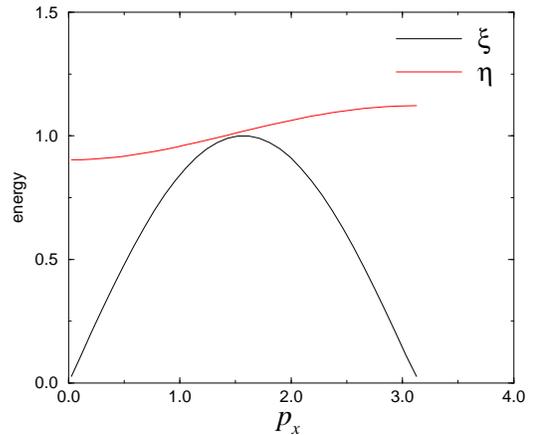}
    \caption{The dispersion relation of $\xi$ and $\eta$ for $m =1000$}
    \label{fig:dispersion2}
  \end{center}
\end{figure}

One might think that larger Majorana mass only results in larger mass
for $\eta$ without any problem, but this is not the case.
Figure~\ref{fig:dispersion2} shows the dispersion relation for large
Majorana mass $m=1000$. 
One can see that although the $\xi$ remains massless, the doublers of
$\xi$ become massless also.

\begin{figure}[htbp]
  \begin{center}
    \leavevmode
    \epsfysize=6cm
    \epsffile{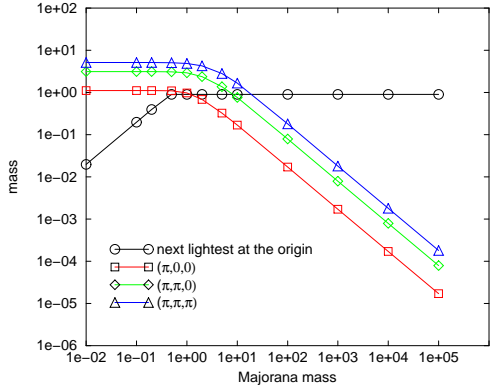}
    \caption{The mass of the next lightest mode at $\bf p = \bf 0$
      as well as that of the doublers of $\xi$ as a function of the
      Majorana mass.} 
    \label{fig:doubler}
  \end{center}
\end{figure}

In order to clarify the situation, we plot in Fig.~\ref{fig:doubler}
the mass of the next lightest mode at $\bf p = \bf 0$ as well as that
of the doublers of $\xi$ as a function of the Majorana mass.
The doublers have mass of the order of the cutoff for $m <1.0$, but
for $m > 1.0$, the mass decreases with increasing Majorana mass as
$\sim 1/m$.
The doublers with many $\pi$'s in the momentum component are heavier
than those with less $\pi$'s.
The mass of the next lightest mode at $\bf p = \bf 0$ grows linearly
as the Majorana mass increases, but saturates for $m>0.5$.
By looking at the wave functions, we find that this mode consists
mostly of $\eta$ for $m < 0.5$, and becomes a mixture of $\xi$ and
$\eta$ for $m > 0.5$.

\section{Interpretation of the result for the large Majorana mass}

The behavior of the doublers for large Majorana mass we have seen
in the previous section can be understood in terms of the seesaw
mechanism.
The situation considered is the case in which Dirac mass and Majorana
mass term coexist.
When we diagonalize the mass matrix of the fermion, a very small
eigenvalue appears when the Majorana mass is much larger than the
Dirac mass.

In fact, the doublers have the two types of mass term in our model.
The Dirac mass term comes from the Wilson term in (\ref{action0}) and
can be written as
\begin{equation}
  S_{\rm Dirac} = 2 n \sum_s ( \bar{\xi}_s \eta_s + \bar{\eta}_s \xi_s
  ),
\end{equation}
where $n$ is the number of $\pi$'s in the momentum component of the
doubler.
Together with the Majorana mass term which comes from
(\ref{majorana}), we have the following mass matrix for the doublers.
\begin{equation}
  \left(
  \begin{array}{cc}
    0 & 2n \\
    2n & m
  \end{array}
  \right).
\end{equation}
The eigenvalues of this matrix for $m \gg n$ are given by $\lambda
\simeq  m$,$\frac{4n^2}{m}$.
The second one is almost zero for $n / m \ll 1$, which explains why
the doublers become very light for large Majorana mass.

\section{Summary}

We examined whether the proposal for decoupling the unwanted zero-mode 
in the domain-wall approach by adding the Majorana mass term for it
works when the gauge field is switched off.
Above all, we clarified what values we should take for the Majorana
mass to be added.
We observed the desired dispersion relation for moderate values of the
Majorana mass, which means that the approach is promising.
We pointed out, however, that for too large Majorana mass, the
doublers of the desired Majorana fermion become very light.
We gave a natural explanation for this phenomenon in terms of the
seesaw mechanism.
We can also obtain the above result by the analysis of the fermion
propagator \cite{aoki}, which will be reported elsewhere \cite{hotta}.
Our next task is of course to switch on the gauge field, which we
would like to report in future publications.

\end{document}